\documentclass[showpacs,a4paper]{revtex4}
\usepackage[english]{babel}
\usepackage[mathscr]{eucal}
\usepackage[dvips]{graphicx}
\usepackage{amsmath}
\usepackage{subfigure}

\def\ds{\displaystyle}
\def\bm#1{\hbox{\boldmath$#1$\unboldmath}}

\def\spacem#1{\setbox1\hbox{$#1$}\hbox to \wd1{\hfill}}

\def\hh{\hrule height0.9pt width1.1em}
\def\vv{\vrule width0.8pt depth0.12em height0.92em}
\def\square{\vbox{\kern0.15em\hh\kern0.9em\hh\kern-1.05em
            \hbox{\vv\kern0.93em\vv}}}
\def\frame#1#2#3#4{\vbox{\hrule height #1pt%
     \hbox{\vrule width #1pt\kern #2pt%
     \vbox{\kern #2pt%
     \hbox{\noindent#4}%
     \kern #2pt}%
     \kern #2pt\vrule width #1pt}%
     \hrule height0pt depth #1pt}}%

\def\sgn{\mathop{\rm sgn}}

\begin{document}
\title{Spin and Pseudospin symmetries in the Dirac equation with central Coulomb potentials}
\author{A. S. de Castro}
\affiliation{Departamento de F{\'\i}sica e Qu{\'\i}mica,
Universidade Estadual Paulista, 12516-410 Guaratinguet\'a, S\~ao
Paulo, Brazil}
\author{P. Alberto}
\affiliation{Physics Department and Centro de F{\'\i}sica
Computacional, University of Coimbra, P-3004-516 Coimbra,
Portugal}
\date{\today}


\begin{abstract}
{We analyze in detail the analytical solutions of the Dirac equation with scalar S and vector V
Coulomb radial potentials near the limit of spin and pseudospin symmetries, i.e.,
when those potentials have the same magnitude and either the same sign or opposite signs,
 respectively. By performing an expansion of the relevant coefficients we also assess the
 perturbative nature of both symmetries and their relations the (pseudo)spin-orbit coupling.
 The former analysis is made for both positive and negative energy solutions and we
 reproduce the relations between spin and pseudospin symmetries found before for nuclear
 mean-field potentials. We discuss the node structure of the radial functions and the quantum numbers
 of the solutions when there is spin or pseudospin symmetry, which we find to be similar to the
 well-known solutions of hydrogenic atoms.}
\end{abstract}

\maketitle

\section{Introduction}

Spin and pseudospin symmetries are $SU(2)$ symmetries of a Dirac Hamiltonian with vector and scalar potentials.
They are realized when the difference, $\Delta=V-S$, or the sum, $\Sigma=V+S$, are constants.
These constants are zero for bound systems whose potentials go to zero at infinity.
Generally, in relativistic quantum physical systems with this kind of potentials neither of these conditions is met exactly but, in some cases, one of them can be approximately true. These symmetries may explain degeneracies in some heavy meson spectra (spin symmetry) or in single-particle energy levels in nuclei (pseudospin symmetry), when these physical systems are described by relativistic mean-field theories with scalar and vector potentials \cite{gino_rev_2005}.
However, for bound systems whose potentials go to zero at infinity, pseudospin symmetry cannot be realized,
because $\Sigma=S+V$ (which must go to zero in the limit of exact pseudospin
symmetry) is also the binding potential. For anti-fermions the opposite occurs, i.e., one cannot have exact
spin symmetry, i.e., $V=S$ \cite{Zhou,ronai_wscc}. This behavior has been related by several authors to the
perturbative nature of spin and pseudospin symmetry \cite{marcos,prc_65_034307,ronai_wscc}.

The problem of a fermion moving in the background of a general mixing of
vector and scalar Coulomb fields has been originally solved for investigating
the implications of a tiny contribution of the scalar Coulomb potential to the atomic spectroscopy \cite{gre1}.
In this paper we explore the fact that this problem has analytical solutions for arbitrary (to a point)
scalar and vector Coulomb radial potentials to examine in a detailed and explicit way, both for fermions and anti-fermions, the onset of spin and pseudospin symmetries and assess their perturbative nature.

This paper is organized as follows. In Sec.~\ref{sec:dirac_coulomb_gen} we
review the general solutions of the Dirac equation with spherical scalar and vector
Coulomb potentials, discuss their quantum numbers and their relation with (pseudo)spin quantum numbers, and
present the formulas for the energy eigenvalues. We also discuss in detail the node structure of the radial wave functions and compare them to what is known from other kind of radial mean-field potentials.
In the next section we present the expansions of the energy eigenvalues for positive and negative energy solutions
in terms of relevant potential parameters for spin and pseudospin symmetries and discuss the perturbative
nature of those symmetries. Next we discuss the quantum numbers and extra degeneracies of the Dirac equation solutions
when there is spin and pseudospin symmetry.
We find that these are similar to the ones found in non-relativistic hydrogenic atoms.
Finally, in Section \ref{sec:conclusions}, we draw the conclusions.

\section{Solutions of the Dirac Hamiltonian with scalar and vector Coulomb potentials}

\label{sec:dirac_coulomb_gen}

The Dirac Hamiltonian with scalar $S$ and vector $V$ potentials reads

\begin{equation}\label{H}
H=\bm\alpha\cdot \bm p\,c + \beta (mc^2 + S) + V\ ,
\end{equation}
where $\bm\alpha$ and $\beta $ are the Dirac matrices in the usual representation
\begin{equation}
\bm\alpha=\left(
\begin{array}{cc}
0 &\bm\sigma \\[1mm]
\bm\sigma & 0
\end{array}
\right), \qquad \beta = \left(
\begin{array}{cc}
I & 0 \\[1mm]
0 & -I
\end{array}
\right)\, ,
\end{equation}
where $\bm\sigma$ are the Pauli matrices and $I$ is the $2\times 2$ unit matrix.
If $S$ and $V$ are of Coulomb type one has
\begin{eqnarray}
 \label{alpha_V} V &=& \frac{\alpha_V}r\,\hbar c \\
  S &=& \frac{\alpha_S}r\,\hbar c
\end{eqnarray}
The Hamiltonian (\ref{H}) can be written in terms of the sum and difference potentials $\Sigma=V+S$ and $\Delta=V-S$ as
\begin{equation}\label{H2}
H=\bm\alpha\cdot \bm p\,c + \beta mc^2 + \frac12(I+\beta)\Sigma + \frac12(I-\beta)\Delta\ ,
\end{equation}
where
\begin{eqnarray}
\label{S_coul}  \Sigma&=& \frac{\alpha_\Sigma} r\,\hbar c \ ,\quad \alpha_\Sigma=\alpha_V+\alpha_S\\
\label{D_coul}  \Delta &=& \frac{\alpha_\Delta}r\,\hbar c \ ,\quad \alpha_\Delta=\alpha_V-\alpha_S
\end{eqnarray}
The solution of the time-independent Dirac equation
\begin{equation}
\label{eq_mov}
H \psi = E \psi \ ,
\end{equation}
with potentials (\ref{S_coul}) and (\ref{D_coul}) is of the form
\begin{equation}
\psi = \left(\begin{array}{c}
\ds i\, \frac{g_{\kappa}(r)}r\, \phi_{\kappa m_j}(\theta,\varphi)\\[3mm]
 \ds-\frac{f_{\tilde\kappa}(r)}r\,\phi_{\tilde\kappa  m_j}(\theta,\varphi)
 \end{array}\right)
\label{psi}
\end{equation}
where
\begin{equation}
\label{def_kappa}
\kappa=\left\{
\begin{array}{cl}
- (\ell+1) & \quad j =  \ell + \frac12 \\[2mm]
   \ell  & \quad j  =  \ell - \frac12
       \end{array}\right.\ ,
\end{equation}
$\ell$ is the orbital angular momentum of the upper component and $\tilde\kappa=-\kappa$.
The angular functions $\phi_{\kappa m_j}(\theta,\varphi)$ are the spinor spherical harmonics
and $g_{n\kappa}(r)$ and $f_{\kappa}(r)$ are the radial
wave functions for the upper and lower components of the Dirac spinor respectively.
The orbital and total angular momenta can be obtained from $\kappa$ by $l=|\kappa|+1/2\big(\kappa/|\kappa|-1\big)$
and $j=|\kappa|-1/2$.

The analytical bound-state solutions of eq. (\ref{eq_mov}) with the Hamiltonian (\ref{H2}) with the potentials
(\ref{S_coul}) and (\ref{D_coul}) can be taken from Greiner and Rafelski book
\cite{greiner_rafelski}, replacing the scalar and vector potentials by their sum and difference.
One has for the radial wave functions, using the notation of Leviatan \cite{Leviatan_prl}
\begin{eqnarray}
\label{g_k}
  g_\kappa(r) &=& -A \sqrt{mc^2+E}\,[(\kappa+\eta_2)\, F_1+n_r F_2]\,\rho^\gamma\,e^{-\rho/2} \\
\label{f_k}
  f_\kappa(r) &=& A \sqrt{mc^2-E}\,[(\kappa+\eta_2)\, F_1-n_r F_2]\, \rho^\gamma\,e^{-\rho/2}
\end{eqnarray}
where $F_1={}_1F_1(-n_r,2\gamma+1,\rho)$ and $F_2={}_1F_1(-n_r+1,2\gamma+1,\rho)$ are confluent hypergeometric
functions and
\begin{eqnarray}
  \rho &=& 2\lambda r \\
  \label{lambda}\lambda &=& \frac1{\hbar c}\sqrt{m^2c^4-E^2} \\
  \label{gamma}\gamma &=& \sqrt{\kappa^2-\alpha_\Delta\alpha_\Sigma} \\
  \label{quantc} n_r&=&-(\gamma+\eta_1)\\
  \label{eta1}  \eta_1&=&\frac1{2\lambda\hbar c}[\alpha_\Sigma(E+mc^2)+\alpha_\Delta(E-mc^2)]\\
  \label{eta2}\eta_2 &=& \frac1{2\lambda\hbar c}[\alpha_\Sigma(E+mc^2)-\alpha_\Delta(E-mc^2)]\ .
\end{eqnarray}
The integers $n_r=0,1,2,\ldots$ are the quantum numbers defined by the quantization condition 
(\ref{quantc}) required in order that the radial functions $g_\kappa(r)$
and $f_\kappa(r)$ be normalizable, and the respective normalization condition
\begin{equation}
\label{norm_cond}
  \int_0^\infty [g^2_\kappa(r)+f^2_\kappa(r)]\,dr=1 \ .
\end{equation}
determines the constant $A$.

One usually defines the principal quantum number as $n=n_r+|\kappa|$ $n=1,2,\ldots$ so that one can write
(\ref{quantc}) as
\begin{equation}
\label{quantc2}
\eta_1=-n+|\kappa|-\gamma=-\xi \ .
\end{equation}
Since $\xi=n_r+\gamma$ is positive definite, this last relation, together with the condition
that $\lambda$, given by (\ref{lambda}), must be real to have bound solutions, yields the
following constraints to the energies and coefficients $\alpha_\Delta$ and $\alpha_\Sigma$:
\begin{eqnarray}
\label{constraint1}
  &&|E|< mc^2  \\
\label{constraint2}
  &&\alpha_\Sigma(E+mc^2)+\alpha_\Delta(E-mc^2) < 0 \ .
\end{eqnarray}
This last equation can be rewritten as
\begin{equation}
\label{constraint3}
\alpha_\Sigma< \alpha_\Delta\frac{mc^2-E}{mc^2+E} \ .
\end{equation}
Another condition comes from the requirement that $\gamma$ be real. One has
\begin{equation}\label{gamma_c}
\alpha_\Sigma\alpha_\Delta<\kappa^2 \ .
\end{equation}
This condition and (\ref{constraint3}) imply that there is no bound-state solution
if $\alpha_\Sigma>0$ and $\alpha_\Delta<0$. If $\alpha_\Sigma$ and $\alpha_\Delta$
have the same sign both conditions (\ref{constraint3}) and (\ref{gamma_c}) 
constrain the values of the strengths of the potentials. When $\alpha_\Sigma<0$ and $\alpha_\Delta>0$
those conditions do not restrict the strength of these potentials.
On the other, since one would also like to have $\alpha_V<0$ (standard
attractive Coulomb potential) for positive energy states, then one must have $|\alpha_\Sigma|>|\alpha_\Delta|$
if the potentials have diferent signs. Thus, in the examples presented below, we shall use values for
the strengths of the Coulomb potentials such that $\alpha_\Sigma<0$, $\alpha_\Delta>0$ and
$-\alpha_\Sigma>\alpha_\Delta$.

From (\ref{quantc}), (\ref{eta1}) and (\ref{eta2}) and  one can derive the following useful relations:
\begin{eqnarray}
\label{eta2^2}
  \eta_2^2&=&\eta_1^2+\alpha_\Delta\alpha_\Sigma \\
\nonumber &=& n_r(n_r+2\gamma)+\kappa^2
\end{eqnarray}

From eqs. (\ref{lambda}), (\ref{eta1}) and (\ref{quantc2}) one can calculate the eigenenergies.
One gets two types of solutions, denoted by $E^{\pm}$:
\begin{equation}
\label{Epm}
E^{\pm}_{n_r,\,\kappa}=mc^2\,\frac{\alpha_\Delta^2-\alpha_\Sigma^2\pm 4\xi\sqrt{\xi^2+\alpha_\Delta\alpha_\Sigma}}
{(\alpha_\Delta+\alpha_\Sigma)^2+4\xi^2}
\end{equation}
The dependence on $n_r$ and $\kappa$ comes through $\xi=n_r+\gamma=n_r+\sqrt{\kappa^2-\alpha_\Delta\alpha_\Sigma}=n-|\kappa|
+\sqrt{\kappa^2-\alpha_\Delta\alpha_\Sigma}$.

The signs were chosen such that $E^+$ is a positive energy state and $E^-$ is a negative energy state when
either $\alpha_\Delta$ or $\alpha_\Sigma$ are very small (see the following section). Since charge conjugation ("c.c.") produces
the changes $V\to -V$ and $S\to S$, $E\to -E$, one has
\begin{equation}
\label{ch_conj}
\begin{array}{l}
  \alpha_\Delta\xrightarrow{\hbox{c.c.}} -\alpha_\Sigma\quad
  \alpha_\Sigma\xrightarrow{\hbox{c.c.}} -\alpha_\Delta\quad
  \xi\xrightarrow{\hbox{c.c.}}\xi\ \hbox{(same quantum numbers $\kappa$ and $n_r$)}\\[2mm]
  E^{\pm}_{n_r,\,\kappa}\xrightarrow{\hbox{c.c.}}E^{\mp}_{n_r,\,\kappa}
\end{array}
\end{equation}

This means that the whole spectrum maps into itself with respect to the charge conjugation operation and therefore contains
 both particle and anti-particle states. Also we note that there is a double degeneracy in the energy levels
 (besides the spin degeneracy), because the eigenenergies (\ref{Epm}) only depend on the magnitude of $\kappa$
 (except when $n_r=0$  --- see the discussion below). As remarked by Leviatan \cite{Leviatan_prl} this degeneracy has
 to do with the fact that the Hamiltonian (\ref{H}), when the scalar and vector potentials are of Coulomb type, commutes with
 the operator
 \begin{eqnarray}
 \label{J}
 J&=&-{\rm i}K\gamma^5(H-\beta mc^2)+\frac{\bm\Sigma\cdot\bm r}{r}(\alpha_V\, mc^2+\alpha_S H) \\
   &=& -{\rm i}K\gamma^5(H-\beta mc^2)+\frac{\bm\Sigma\cdot\bm r}{r}[(\alpha_\Sigma+\alpha_\Delta) mc^2+(\alpha_\Sigma-\alpha_\Delta) H]/2\ .
 \end{eqnarray}
where
\begin{equation}
\bm\Sigma=\left(
\begin{array}{cc}
\bm\sigma &0 \\[1mm]
0 & \bm\sigma
\end{array}
\right)\
\end{equation}
and $K$ is related to the $\bm L\cdot \bm S$ operator (see section \ref{sec_spin_psin}) such that
the spinor (\ref{psi}) is its eigenspinor with eigenvalue $\kappa$. This is a generalization of the Johnson-Lippmann operator
\cite{Leviatan_prl,Johnson-Lippmann}, which is itself a generalization of the Laplace-Runge-Lenz operator of the
non-relativistic quantum Coulomb problem to the relativistic one.
Since $J$ anti-commutes with $K$, the energy must depend only on $|\kappa|$.

 One may check that the non-relativistic limit is correct. If one sets $\alpha_\Sigma=\alpha_\Delta=\alpha_V$
 and have $\alpha_V\ll\xi\,,\xi\sim n$, one gets
 \begin{equation}\label{non-rel-lim}
 E^+-mc^2\sim -mc^2\frac{\alpha_V^2}{2n^2}
 \end{equation}
 Given that, from (\ref{alpha_V}), we may set, for the quantum system of an electron in a atom of atomic number $Z$,
 $\alpha_V=-Z\alpha$,
 where $\alpha=e^2/(\hbar c)$ is the fine structure constant, equation (\ref{non-rel-lim}) is just the energy of
 the level $n$ of an electron of a hydrogenic atom with atomic number $Z$.

\subsection{Node structure of the solutions}

We discuss now the node structure of the radial wave functions (\ref{g_k}) and (\ref{f_k}).
The solution with $n_r=0$ deserves a special consideration. In this case the radial functions
$g_\kappa(r)$ and $f_\kappa(r)$ will reduce to
\begin{eqnarray}
\label{g_nr=0}
  g_\kappa(r) &=& -B \sqrt{mc^2+E}\,(\kappa+\eta_2)\,\rho^\gamma\,e^{-\rho/2} \\
\label{f_nr=0}
  f_\kappa(r) &=& B \sqrt{mc^2-E}\,(\kappa+\eta_2)\,\rho^\gamma\,e^{-\rho/2}
\end{eqnarray}
where $B$ is a constant.
These functions have only one node at the origin. Furthermore, we may note that,
if $\alpha_\Sigma$ and $\alpha_\Delta$ have the same sign, $\gamma<1$, and therefore
the radial wave functions $g_\kappa(r)/r$ $f_\kappa(r)/r$, although
integrable, would be singular at the origin.
Another important point is that, in this case,
from (\ref{eta2^2}), $\eta_2=\pm|\kappa|$. Because of the factor $\kappa+\eta_2$ in
eqs.~(\ref{g_nr=0}) and (\ref{f_nr=0}) a non-zero wave function
cannot exist for any sign of $\kappa$. It turns out, as already remarked by Leviatan
\cite{Leviatan_prl}, that for $E^+$ solutions one must have $\kappa<0$ while for
$E^-$ solutions one must have $\kappa>0$, which means that for $n_r=0$ the degeneracy
in the sign of $\kappa$ is broken.

For $n_r\ge 1$, $g_\kappa(r)$ and $f_\kappa(r)$ can be written as
\begin{eqnarray}
\label{g_k2_Lg}
  g_\kappa(r) &=& -A' \sqrt{mc^2+E}\,[(\kappa+\eta_2)\, L_{n_r}^{2\gamma}(\rho)+(2\gamma+n_r)L_{n_r-1}^{2\gamma}(\rho)]\,\rho^\gamma\,e^{-\rho/2} \\
\label{f_k2_Lg}
  f_\kappa(r) &=& A' \sqrt{mc^2-E}\,[(\kappa+\eta_2)L_{n_r}^{2\gamma}(\rho)-
  (2\gamma+n_r)L_{n_r-1}^{2\gamma}(\rho)]\, \rho^\gamma\,e^{-\rho/2}
\end{eqnarray}
where $L_{n_r}^{2\gamma}(\rho)$ are the generalized Laguerre polynomials of degree $n_r=n-|\kappa|$ and $A'$ is a constant.
These polynomials have $n_r$ distinct positive zeros.
%except possibly, as remarked above, if $\alpha_\Sigma$ and $\alpha_\Delta$ have the same sign.
These formulas suggest, then, that $g_\kappa(r)$ and
$f_\kappa(r)$ may have the same number of nodes, namely $n_r+1$ ($n_r=n-|\kappa|$ nodes for $r>0$ plus one at $r=0$).
In the following we prove that, if $\alpha_\Sigma\alpha_\Delta<0$ then indeed the number of nodes of $g_\kappa(r)$, $n_g$,
and the number of nodes of $f_\kappa(r)$, $n_f$, are both equal to $n_r+1$ for any sign of $\kappa$.

The zeros of generalized Laguerre polynomials of different degrees are interweaved, i.e., the $i$th zero of
$L_{n_r-1}^{2\gamma}(\rho)$ is localized between the $i$th and $(i+1)$th zero of $L_{n_r}^{2\gamma}(\rho)$ and their value
at $\rho=0$ is positive and increases with the polynomial degree. As a consequence, one finds that the number of zeros $n_L$ of the linear combination ($A\not=0$, $B\not=0$)
\begin{equation}
\label{L_rho}
    L(\rho)=A\,L_{n_r}^{2\gamma}(\rho)+B\,L_{n_r-1}^{2\gamma}(\rho)
\end{equation}
depend on the ratio $B/A$ in the following way
\begin{equation}
\label{zeros_L}
n_L=\left\{
\begin{array}{cl}
n_r-1 & ,\quad\ds\frac{B}{A}<-\frac{L_{n_r}^{2\gamma}(0)}{L_{n_r-1}^{2\gamma}(0)} \\[4mm]
n_r & ,\quad\ds\frac{B}{A}\geq -\frac{L_{n_r}^{2\gamma}(0)}{L_{n_r-1}^{2\gamma}(0)}
\end{array}\right.\ .
\end{equation}
From the values at the origin of associated Laguerre polynomials (see, for instance,
\cite{abramowitz}) one has
\begin{equation}
\label{ratio_L(0)}
\frac{L_{n_r}^{2\gamma}(0)}{L_{n_r-1}^{2\gamma}(0)}=\frac{n_r+2\gamma}{n_r}\ .
\end{equation}
Since there is an extra node at $r=\rho=0$ and that the equality in (\ref{zeros_L}) corresponds 
to a node at the origin, from eqs.~(\ref{g_k2_Lg}) and (\ref{f_k2_Lg}) one may write 
\begin{eqnarray}
\nonumber
n_g &=& \left\{
\begin{array}{ll}
n_r & ,\quad\ds\frac{2\gamma+n_r}{\kappa+\eta_2}\leq-\frac{n_r+2\gamma}{n_r} \\[4mm]
n_r+1 & ,\quad\ds\frac{2\gamma+n_r}{\kappa+\eta_2}>-\frac{n_r+2\gamma}{n_r}
\end{array}
\right.\ ,\\
\nonumber
n_f &=& \left\{
\begin{array}{ll}
n_r & ,\quad\ds-\frac{2\gamma+n_r}{\kappa+\eta_2}\leq-\frac{n_r+2\gamma}{n_r} \\[4mm]
n_r+1 & ,\quad\ds-\frac{2\gamma+n_r}{\kappa+\eta_2}>-\frac{n_r+2\gamma}{n_r}
\end{array}\right.\ .
\end{eqnarray}
This can be further simplified to get
\begin{eqnarray}
\label{n_g2}
n_g &=& \left\{
\begin{array}{ll}
n_r & ,\quad |\kappa+\eta_2|\leq -\sgn(\kappa+\eta_2) n_r \\[4mm]
n_r+1 & ,\quad|\kappa+\eta_2|> -\sgn(\kappa+\eta_2) n_r
\end{array}
\right.\ ,\\
\label{n_f2}
n_f &=& \left\{
\begin{array}{ll}
n_r & ,\quad |\kappa+\eta_2|\leq \sgn(\kappa+\eta_2) n_r \\[4mm]
n_r+1 & ,\quad|\kappa+\eta_2|> \sgn(\kappa+\eta_2) n_r
\end{array}\right.\ ,
\end{eqnarray}
where $\sgn(x)$ is the signum function giving the sign of $x$.

On the other hand, from (\ref{gamma}) and (\ref{eta2^2}) and when $\alpha_\Sigma\alpha_\Delta<0$,
the following inequalities hold
\begin{eqnarray}
\nonumber
 |\eta_2|& > & n_r \\
\nonumber
 \gamma & > & |\kappa| \Rightarrow |\eta_2|> n_r+|\kappa|
\end{eqnarray}
meaning that
\begin{equation}
|\kappa+\eta_2| > n_r
\end{equation}
which is true when $\kappa$ and $\eta_2$ have the either the same or different signs, which is to say
that it is true for \textit{any} sign of $\kappa$ and $\eta_2$. From (\ref{n_g2}) and
(\ref{n_f2}) it follows that, when $\alpha_\Sigma\alpha_\Delta<0$, one has
\begin{eqnarray}
\label{n_g3}
n_g &=& n_r+1 \\
\label{n_f3}
n_f &=& n_r+1
\end{eqnarray}
for \textit{any} $\kappa$.

Note that this result applies to both $E^+$ and $E^-$ solutions. Also, since
$\alpha_\Sigma$ and $\alpha_\Delta$ have different signs, $\gamma > 1$ and thus
$g_\kappa(r)/r$ and $f_\kappa(r)/r$ have the same nodes as $g_\kappa(r)$ and $f_\kappa(r)$.

\begin{figure}[!ht]
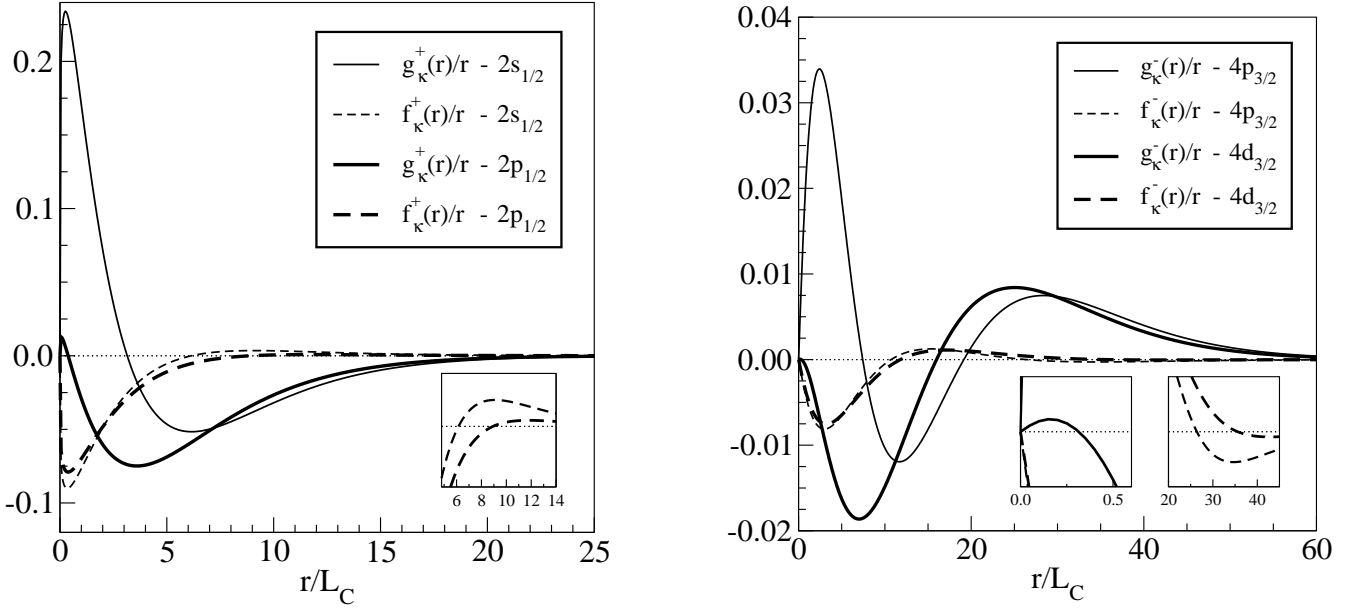

{
\parbox[!ht]{8.2cm}{
\begin{center}
\includegraphics[width=8cm]{fig1a.eps}
\end{center}
}\hfill
\parbox[!ht]{8.2cm}{
\begin{center}
\includegraphics[width=8cm]{fig1b.eps}
\end{center} }  } \vspace*{-.2cm}
 \caption{Radial wave functions $g_\kappa(r)/r$ and $f_\kappa(r)/r$ for
 the $E^+$ states $2s_{1/2}$, $2p_{1/2}$ ($\kappa=\mp 1$, $n_r=1$, $n_{g,f}=2$) and
 the $E^-$ states $4p_{3/2}$, $4d_{3/2}$ ($\kappa=\mp 2$, $n_r=2$, $n_{g,f}=3$),
 when $\alpha_\Delta=0.5$ and $\alpha_\Sigma=-0.8$.
 $L_C=\hbar/(mc)$ is the Compton wavelength.\hfill\ } \label{fig:F_G_2s_4p}
\end{figure}

In Fig.~\ref{fig:F_G_2s_4p} this is shown explicitly for the $E^+$ levels $2s_{1/2},\ 2p_{1/2}$ ($\kappa=\mp 1$, $n_r=1$) (left panel) and for the $E^-$ levels $4p_{3/2},\ 4d_{3/2}$ ($\kappa=\mp 2$, $n_r=2$) (right panel), for $\alpha_\Delta=0.5$ and $\alpha_\Sigma=-0.8$.
The radial coordinate is given in units of the Compton wavelength $L_C=\hbar/(mc)$. Here one uses, as usual,
the spectroscopic notation $n\ell_j$ referring to the upper component of the Dirac spinor.

It is interesting to remark that this node structure does not follow the rule for
the Dirac equation with vector and scalar radial potentials derived by Leviatan and Ginocchio
in \cite{levi_gino},
\begin{eqnarray}
n_{f}=\left\{
\begin{array}{ll}
n_{g} &,\;\ \ \ \kappa<0\\[0.2cm]
n_{g}+1   &,\;\ \ \ \kappa>0
\end{array}
\right.\ ,\label{Eq:nodes}
\end{eqnarray}
by which only for negative $\kappa$ the upper radial and lower radial functions
of the Dirac spinor have the same number of nodes. The structure (\ref{Eq:nodes}) also holds
for anti-particles in Woods-Saxon scalar and vector mean-field potentials, as shown
in \cite{ronai_wscc}. This may be because these potentials were finite at $r=0$, as are the nuclear mean-field scalar and vector potentials, and thus the behavior of the respective radial functions at the origin is different form the present case of Coulomb potentials.

\section{Spin and pseudospin symmetries in the Dirac Hamiltonian with scalar and vector Coulomb potentials}
\label{sec_spin_psin}

To study the conditions under which the Dirac Hamiltonian
(\ref{H2}) with the Coulomb potentials (\ref{S_coul}) and (\ref{D_coul}) has bound solutions
near the conditions of spin and pseudospin symmetry, we will perform an expansion of
eq. (\ref{Epm}) for small $\alpha_\Delta$ (spin) and $\alpha_\Sigma$ (pseudospin).

If we expand (\ref{Epm}) (using the principal quantum number $n$) until next-to-next-to-leading order in $\alpha_\Delta$ one gets
{\setlength\arraycolsep{0pt}
\begin{eqnarray}
\label{epdelta}
\nonumber
E^{+}_{n,\,\kappa} = mc^2 \Bigg\{&& 1-\frac{2 \alpha_\Sigma^2}{\alpha_\Sigma^2+4 n^2}+\frac{4
   \alpha_\Sigma^3 \left(\kappa^2-2 n |\kappa|\right)}{\kappa^2 \left(\alpha_\Sigma^2+4
   n^2\right)^2}\;\alpha_\Delta\\
 && \left.-\frac{\alpha_\Sigma^4 \left\{4 n^3 |\kappa|
   \left[\alpha_\Sigma^2+4 \left(n^2-4 \kappa^2\right)\right]+\kappa^2
   \left[\alpha_\Sigma^2 \left(\kappa^2-4 n^2\right)+20 \kappa^2 n^2+48
   n^4\right]\right\}}{2 \kappa^4 n^2 \left(\alpha_\Sigma^2+4
   n^2\right)^3}\;\alpha_\Delta^2+{\cal O}\left(\alpha_\Delta^3\right)\right\}\\
\label{emdelta}
E^{-}_{n,\,\kappa}=mc^2\Bigg[&&-1+\frac{\alpha_\Delta^2}{2 n^2}-\frac{\alpha_\Sigma
   \left(\kappa^2-2 n |\kappa|\right)}{4 \kappa^2
   n^4}\;\alpha_\Delta^3 +{\cal O}\left(\alpha_\Delta^4\right)\Bigg]\ ,
   \end{eqnarray}
}
while for $\alpha_\Sigma$ one has
{\setlength\arraycolsep{0pt}
\begin{eqnarray}
\label{epsigma}
E^{+}_{n,\,\kappa} = mc^2 \Bigg[&&1-\frac{\alpha_\Sigma^2}{2 n^2}+
\frac{\alpha_\Delta  \left(\kappa^2-2 n
   |\kappa|\right)}{4 \kappa^2 n^4}\;\alpha_\Sigma^3+{\cal O}\left(\alpha_\Sigma^4\right)\Bigg]
 \\
\label{emsigma}
\nonumber
E^{-}_{n,\,\kappa}=mc^2\Bigg\{&&-1+
   \frac{2 \alpha_\Delta^2}{\alpha_\Delta^2+4 n^2}-\frac{4
   \alpha_\Delta^3 \left(\kappa^2-2 n |\kappa|\right)}{\kappa^2
   \left(\alpha_\Delta^2+4 n^2\right)^2}\;\alpha_\Sigma\\
&&   +\frac{\alpha_\Delta^4
   \left\{4 n^3 |\kappa| \left[\alpha_\Delta^2+4 \left(n^2-4 \kappa^2\right)\right]+\kappa^2
   \left[\alpha_\Delta^2 \left(\kappa^2-4 n^2\right)+20 \kappa^2 n^2+48
   n^4\right]\right\}}{2 \kappa^4 n^2 \left(\alpha_\Delta^2+4
   n^2\right)^3}\;\alpha_\Sigma^2+{\cal O}\left(\alpha_\Sigma^3\right)\Bigg\}
   \end{eqnarray}
}

From these equations is clear that $E^{+(-)}$ are the energies of bound positive (negative)
energy states for small $\alpha_\Delta$ and $\alpha_\Sigma$. The symmetry of the expressions
above when one switches from $\alpha_\Delta$ to $\alpha_\Sigma$ and from positive to negative
energy states is just the illustration of the charge conjugation transformations
as given by eqs.~(\ref{ch_conj}).
From the equations above one sees also that when the zeroth-order term is just $\pm mc^2$
the respective symmetry cannot be realized, since there is no bound
states. For positive energy states one sees that in eq.~(\ref{epsigma}) (pseudo-spin symmetry)
and for negative energy states in eq.~(\ref{emdelta}) (spin symmetry).
Again, this agrees with what is known from realistic calculations in nuclei, since $\Sigma$
and $-\Delta$ potentials act as biding potentials for fermions and antifermions respectively
\cite{Zhou,ronai_wscc}.

It is interesting to examine how the spin and pseudospin symmetries get broken and the corresponding
energy splittings of the (pseudo)spin partners.

%To that end we must, as usual, label the energy states according
%to the nodes of the upper radial function $g_\kappa$. This function can be written as
%\begin{equation}
%% \nonumber to remove numbering (before each equation)
%  g_\kappa(r) =-B\sqrt{mc^2+E}\,[(\kappa+\eta_2)\, L_{n_r}^{2\gamma}(\rho)+(2\gamma+n_r)\,L_{n_r-1}^{2\gamma}(\rho)]\,\rho^\gamma\,e^{-\rho/2}
%\end{equation}
%where $L_{n}^{\alpha}(\rho)$ are associated Laguerre polynomials.

The next to leading order terms of equations (\ref{epdelta}-\ref{emsigma}) reveal that the breaking of spin symmetry
is perturbative and the breaking of pseudosymmetry is non-perturbative for $E^+$ states.
The reverse is true for $E^-$ states. Indeed, this can be seen the fact that the next to leading order
terms in eqs.~(\ref{epdelta}) and (\ref{epsigma}) are first-order in $\alpha_\Delta$ and $\alpha_\Sigma$,
respectively. This is what we expect from first-order perturbation theory and also from the fact that
the breaking of spin and pseudo-spin symmetries from their exact realizations should come from the
spin-orbit and pseudospin-orbit coupling terms, which can be obtained from the second-order radial equations for
$g_\kappa(r)$ and $f_\kappa(r)$, respectively (see, for instance, ref.~\cite{prc_65_034307})

{\setlength\arraycolsep{1.4pt}
\begin{eqnarray}
\nonumber
(\hbar c)^2\biggl[\frac{1}{r^2}\frac{d\,}{dr}\bigg(r^2\frac{d\,}{dr}\bigg)-\frac{\kappa (\kappa+1)%
}{r^{2}}+\frac{\Delta ^{\prime }}{{E}+mc^2-\Delta (r)}\biggl(\frac{%
d\,\,}{dr}&+&\frac{1+\kappa}{r}\biggr)\biggr]\frac{g_{\kappa }(r)}r\\
 &=&-[{E}-mc^2-\Sigma (r)][E+mc^2-\Delta (r)]\frac{g_{\kappa }(r)}r\ ,
\label{Eq:D2ordgGeral}\\
\nonumber
(\hbar c)^2\biggl[\frac{1}{r^2}\frac{d\,}{dr}\bigg(r^2\frac{d\,}{dr}\bigg) -\frac{\kappa (\kappa -1)%
}{r^{2}}+\frac{\Sigma ^{\prime }}{{E}-mc^2-\Sigma (r)}\biggl(\frac{%
d\,}{dr}&+&\frac{1-\kappa }{r}\biggr)\biggr]\frac{f_{\kappa }(r)}r\\
&=&-[E-mc^2-\Sigma (r)][E+mc^2-\Delta (r)]\frac{f_{\kappa }(r)}r\ ,
\label{Eq:D2ordfGeral}
\end{eqnarray}
}
%-----------------------------------------------------------------------------
where the prime means derivative with respect to $r$.
%-------------------------------------------------------------------------------
These are
proportional to the derivatives of $\Delta$ (spin) and $\Sigma$ (pseudospin) and thus, for small
enough values of $\alpha_\Delta$ and $\alpha_\Sigma$, respectively,
their expectation values contributing to the energy splitting of spin and pseudospin partners
should be linear in those parameters. In particular, one may relate the
contributions from spin-orbit and pseudo-orbit couplings to the total energy
to the following terms coming coming from eqs. (\ref{Eq:D2ordgGeral})
and (\ref{Eq:D2ordfGeral}) as was done in ref. \cite{prc_65_034307}:

{\setlength\arraycolsep{1.4pt}
\begin{eqnarray}
\label{E_spin-orbit}
 E^{\rm SO}&=& - (\hbar c)^2 \ds\int_0^\infty\frac{\Delta ^{\prime }}{(E+mc^2-\Delta)^2}\frac{1+\kappa }{r}\,
  g_{\kappa}^2\,{\rm d}r\bigg/\bigg(\int_0^\infty  g_{\kappa}^2\,{\rm d}r\bigg) \\
\label{E_pspin-orbit}
  E^{\rm PSO}&=& -(\hbar c)^2 \ds\int_0^\infty\frac{\Sigma ^{\prime }}
  {(E-mc^2-\Sigma)(E+mc^2-\Delta)}%
\frac{1-\kappa }{r}\,
 f_{\kappa}^2\,{\rm d}r \bigg/\bigg(\int_0^\infty  f_{\kappa}^2\,{\rm d}r\bigg)\ .
\end{eqnarray}
}
These terms are consistent with the non-relativistic formula for spin-orbit coupling, the first one applied to positive energy states and
the second one to negative energy states. The origin of the factors
$1+\kappa$ for the upper component and to $1-\kappa$ for the lower component can be traced to the action of the operator
$\bm L\cdot \bm S$ upon the spinor (\ref{psi}), since $\bm L\cdot \bm S\,\psi=
-\hbar^2/2(\beta K+I)\psi$ with $K\psi=\kappa\psi$ and $\bm S=\hbar/2\,\bm\Sigma$.
However, contrary to what happens with Woods-Saxon type potentials
for positive energy states \cite{prc_65_034307}, the spin-orbit term (\ref{E_spin-orbit}) has a double pole at $r=\hbar c\,\alpha_\Delta/(E+mc^2)$
so it cannot be calculated separately from the so-called Darwin term, coming from the derivative term in
eq.~(\ref{Eq:D2ordgGeral}) and also multiplied to $\Delta ^{\prime }$, since the sum of the two terms must of course be finite.

Therefore, for vector and scalar Coulomb
potentials, one can establish a connection between the realization of exact pseudospin and spin symmetries for bound states and their perturbative breaking. This was also recently shown to be the case for vector and scalar
Woods-Saxon nuclear mean-field potentials for neutrons and anti-neutrons \cite{ronai_wscc}.

\subsection{Quantum numbers for exact spin and pseudospin symmetries}

Finally, we discuss the quantum numbers of states when there is either exact spin or pseudospin symmetries, i.e., when
$\alpha_\Delta=0$ or $\alpha_\Sigma=0$ respectively.
The energies of those levels are given by

%{\setlength\arraycolsep{0.4ex}
%\begin{eqnarray}
%% \nonumber to remove numbering (before each equation)
%E^{+}_{n,\,\kappa} &=& mc^2 \bigg( 1-\frac{2 \alpha_\Sigma^2}{\alpha_\Sigma^2+4 n^2}\bigg)
%\qquad({\rm spin}) \\
%E^{-}_{n,\,\kappa}&=&mc^2\bigg(-1+
%   \frac{2 \alpha_\Delta^2}{\alpha_\Delta^2+4 n^2}\bigg)\qquad({\rm pseudospin})
%\end{eqnarray}
%}
\begin{eqnarray}
\label{E_spin_sym}
 E^{+}_{n,\,\kappa} &=& mc^2 \bigg( 1-\frac{2 \alpha_\Sigma^2}{\alpha_\Sigma^2+4 n^2}\bigg)  \qquad({\rm spin}) \\[2mm]
\label{E_pspin_sym}
  E^{-}_{n,\,\kappa}&=&mc^2\bigg(-1+
   \frac{2 \alpha_\Delta^2}{\alpha_\Delta^2+4 n^2}\bigg)\qquad({\rm pseudospin})
\end{eqnarray}

One may notice immediately that these energies only depend on the principal quantum number $n$.
This situation looks very similar to what happens with the energy levels of a non-relativistic hydrogenic atom.
To see how it comes about we start out with spin symmetry case ($\Delta=0$ or $\alpha_\Delta=0$).
The corresponding Dirac equation reads
\begin{equation}
\label{Dirac_Delta=0}
[\bm\alpha\cdot \bm p\,c + \beta mc^2 + \frac12(I+\beta)\Sigma]\,\psi=E\psi\ ,
\end{equation}
with $\Sigma=\alpha_\Sigma\hbar c/r$.

Following the procedure of \cite{prc_75_047303} we use the projectors
$\psi_\pm=P_\pm\psi=[(I\pm\beta)/2]\,\psi$ such that
\begin{equation}
\psi_+=\left(\begin{array}{c}
\phi\\[2mm] 0
\end{array}\right)\qquad \psi_-=\left(\begin{array}{c}
0\\[2mm] \chi
\end{array}\right) \ ,
\end{equation}
where $\phi$ and $\chi$ are respectively the upper and lower
two-component spinors. Using the properties and anti-commutation
relations of the matrices $\beta$ and $\bm\alpha$ we can apply
the projectors $P_\pm$ to the Dirac equation (\ref{Dirac_Delta=0}) and
decompose it into two coupled equations for $\psi_+$ and $\psi_-$:
\begin{eqnarray}
\label{psi-1}
  c\,\bm\alpha\cdot\bm p\,\psi_-+(mc^2 + \Sigma)\,\psi_+&=&E\psi_+ \\
\label{psi+1}
 c\,\bm\alpha\cdot\bm
 p\,\psi_+-mc^2 \,\psi_-&=&E\psi_- \ .
\end{eqnarray}
From this last equation we get $\psi_-$ and replace it in (\ref{psi-1}) to give
\begin{equation}
\label{Dirac_psi+_Delta=0}
\bm p^2\,c^2\,\psi_+=(E+mc^2)(E-mc^2-\Sigma)\psi_+\ .
\end{equation}
or yet
\begin{equation}
\label{Dirac_psi+_Delta=0_2}
\frac{\bm p^2}{2m}\,\psi_+=(\mathcal{E}'-\Sigma')\,\psi_+\ ,
\end{equation}
where
\begin{equation}
\label{definitions}
\mathcal{E}'=\bigg(\frac{\mathcal{E}}{2mc^2}+1\bigg)\mathcal{E}\ ,\qquad \Sigma'=\bigg(\frac{\mathcal{E}}{2mc^2}+1\bigg)\Sigma\ ,\qquad
\mathcal{E}=E-mc^2 \ .
\end{equation}
Equation (\ref{Dirac_psi+_Delta=0_2}) is just the Schroedinger equation for a hydrogenic atom of ``atomic number"
\begin{equation}
Z'=-\bigg(\frac{\mathcal{E}}{2mc^2}+1\bigg)\alpha_\Sigma/\alpha\ ,
\end{equation}
where again $\alpha$ is the fine structure constant. The energy of this ``atom" is just (see eq.~(\ref{non-rel-lim}))
\begin{equation}
\mathcal{E}'=-mc^2\bigg(\frac{\mathcal{E}}{2mc^2}+1\bigg)^2\frac{\alpha_\Sigma^2}{n^2}
\end{equation}
Replacing the expressions (\ref{definitions}) into this last equation one gets eq.~(\ref{E_spin_sym}).

The fact that the energy only depends on $n$ in spin symmetry conditions implies also that there is
an extra level degeneracy. Because of the relation $n=n_r+|\kappa|$, levels with the same principal
quantum number and $|\kappa|\leq n$
are degenerate. This also has a correspondance to the well known degeneracy of levels of
a hydrogenic atom in which all levels with $\ell\leq n-1$ and the same $n$ are degenerate.
These properties can be traced back to the fact that in spin symmetry conditions and radial potentials
there is an SU(2) symmetry generated by the operator
\begin{equation}
\label{cal_L}
 \bm{\mathcal{L}}=\bm L
P_++\frac1{p^2}\,\bm\alpha\cdot\bm p\,\bm
L\,\bm\alpha\cdot\bm p\,P_-=
\left(\begin{array}{cc} \bm L&0\\[2mm]
0&U_p\,\bm L\,U_p\end{array}\right)\ ,
\end{equation}
where $U_p=\bm\sigma\cdot\bm p/(\sqrt{p^2})$ is the helicity
operator. This means that the orbital angular momentum $\ell$ of the upper component of
the Dirac spinor is a good quantum number is this case \cite{gino_rev_2005,prc_75_047303}.
Another way of stating this is
saying that when there is spin symmetry there is no spin-orbit coupling.

A similar reasoning can be made when there is pseudospin symmetry ($\Sigma=0$, ).
In this case one would get the following equation for the lower component $\psi_-$:
\begin{equation}
\label{Dirac_psi-_Sigma=0}
\frac{\bm p^2}{2m}\,\psi_-=(\mathcal{E}'-\Delta')\,\psi_-\ ,
\end{equation}
where
\begin{equation}
\label{definitions_2}
\mathcal{E}'=\bigg(\frac{\mathcal{E}}{2mc^2}+1\bigg)\mathcal{E}\ ,\qquad \Delta'=-\bigg(\frac{\mathcal{E}}{2mc^2}+1\bigg)\Delta\ ,\qquad
\mathcal{E}=-E-mc^2 \ .
\end{equation}

As one would expect for negative energy states and from charge conjugation, this amounts basically to change the energy sign and
also replacing $\Sigma$ by $-\Delta$ \cite{ronai_wscc}. The SU(2) generator would be in this case
\begin{equation}
\tilde{\bm{\mathcal{L}}}=\frac1{p^2}\,\bm\alpha\cdot
\bm p\,\bm L\,\bm\alpha\cdot\bm p\,P_++\bm L P_-=
\left(\begin{array}{cc} U_p\,\bm L\,U_p&0\\[2mm]
0&\bm L\end{array}\right)\ ,
\end{equation}
and in this case the orbital angular momentum $\tilde\ell$ of the lower component of
the Dirac spinor is a good quantum number \cite{gino_rev_2005,prc_75_047303}.

\section{Conclusions}
\label{sec:conclusions}

We analyzed in detail the analytical solutions of Dirac equation with scalar S and vector V
Coulomb radial potentials in terms of the sum $\Sigma$ and difference $\Delta$ potentials.
Besides reviewing the solutions at this light, we established the node structure of their radial solutions
and examined the solutions near the limit of spin and pseudospin symmetries, i.e.,
when when $\Delta=0$ and $\Sigma=0$ respectively. We assessed the perturbative nature of those symmetries,
and confirmed previous results with another type of potentials, namely, that spin symmetry is perturbative and
pseudospin symmetry in non-perturbative for positive energy solutions, while the reverse happens for negative
energy solutions. We also found that in conditions of exact spin and pseudospin symmetries the solutions of
the Dirac equation have similar features of the well-known non-relativistic solutions for
hydrogenic atoms, namely that they depend only on the principal quantum number and exhibit a similar
 degeneracy.

\begin{acknowledgments}
We acknowledge financial support from CNPQ and QREN/FEDER,
the Programme COMPETE, under Project No.~PTDC/FIS/113292/2009.
P. Alberto would like also to thank the Universidade Estadual
Paulista, Guaratinguet\'a campus, for supporting his stays in its
Physics and Chemistry Department.
\end{acknowledgments}

\end{document}